\documentclass[12pt]{article}
\usepackage{amssymb,amscd,amsmath,amsthm}
\usepackage{latexsym,amstext}
\usepackage{latexsym,amstext}
\usepackage{color}
\usepackage{graphicx}
\usepackage{epstopdf}
\usepackage{cite}

\usepackage{amssymb,amscd,amsmath,amsthm}
\usepackage{latexsym,amstext}
\usepackage{dcolumn}
\usepackage{color}

\def\be{\begin{equation}}
\def\ee{\end{equation}}
\def\bea{\begin{eqnarray}}
\def\eea{\end{eqnarray}}
\def\bean{\begin{eqnarray*}}
\def\eean{\end{eqnarray*}}

\def\e{{\mathcal E}}
\def\I{{\mathbb I}}
\def\K{{\mathcal K}}
\def\N{{\mathbb N}}
\def\Z{{\mathbb Z}}
\def\R{{\mathbb R}}
\def\C{{\mathbb C}}
\def\a{{\alpha}}
\def\ds{\displaystyle}

\begin{document}

\title{\bf Hermite functions, Lie groups and  Fourier Analysis}

\author{E. Celeghini$^{1,2}$, M. Gadella$^{2,3}$, M. A. del Olmo$^{2,3}$\\ \\
$^1$ Dipartimento di Fisica, Universit\`a di Firenze and\\  INFN-Sezione di Firenze\\
150019
Sesto Fiorentino, Firenze, Italy\\
$^2$ Departamento de F\'{\i}sica Te\'orica, At\'omica y Optica  and IMUVA, \\
Universidad de Va\-lladolid, 47011 Valladolid, Spain\\ 
$^3$ Instituto de Matem\'aticas, Universidad de Valladolid (IMUVA) \\ \\
celeghini@fi.infn.it\\
manuelgadella1@gmail.com\\
 marianoantonio.olmo@uva.es\\  \\
 Keywords: \em Fourier analysis; Special functions; Rigged Hilbert spaces; \\ \em Quantum mechanics; Signal proceessing
 }

\maketitle

\begin{abstract}
In this paper, we present recent results in harmonic analysis in the real line $\R$ and in the half-line $\R^+$, which show a closed relation between Hermite and Laguerre functions, respectively, their symmetry groups and Fourier analysis.  This can be done in terms of a unified framework based in the use of rigged Hilbert spaces. We find a relation between the universal enveloping algebra of the symmetry groups with the fractional Fourier transform. The results obtained are relevant in quantum mechanics as well as in signal processing as Fourier analysis has a close relation with signal filters. Also, we introduce some new results concerning a discretized Fourier transform on the circle. We introduce new functions on the circle constructed with the use of Hermite functions with interesting properties under Fourier transformations.
\end{abstract}

\section{Introduction}

The seminal work by Fourier of 1807, published in 1822 \cite{fourier}, about the solution of the heat equation had a deep impact in physics and mathematics as is well known. Roughly speaking, the Fourier method decomposes functions into a superposition of ``special functions'' \cite{berry,andrews}. Among these special functions, trigonometric functions, originally used by Fourier himself, play a fundamental role in the standard applications of Fourier series as well as Fourier transforms. In addition, the Fourier method makes use of other types of special functions, which in many cases are associated with groups. In this latter case, these special functions have symmetry properties, inherited  from the corresponding group, and this is for instance the way in which harmonic analysis appears in group representation theory \cite{folland15}. An interesting aspect of Fourier analysis is the decomposition of Hilbert space vectors, quite often square integrable functions on some domain, into orthogonal basis. This generalizes both, the standard Fourier analysis of trigonometric series and the decomposition of a vector in terms of an algebraic basis of linearly independent vectors. Another generalization is the decomposition of a self-adjoint or normal operator on a Hilbert space in terms of spectral measures, say through the spectral representation theorem. We are mainly interested in these generalizations concerning Hilbert space vectors and operators.

In a recent work \cite{CGO16} we have revisited the  harmonic analysis on the real line and  studied the relation between the Hermite functions, the Weyl-Heisenberg Lie algebra and the Fourier analysis inside the framework of rigged Hilbert spaces. As is well known, the Fourier transform relates two continuous bases which are used in the description of 
one-dimensional quantum systems on the real line. These are the coordinate and momentum representations, naturally connected with the position and the momentum operator \cite{CO,FO}, respectively. They span the Weyl-Heisenberg algebra  together with the identity operator. Moreover these two continuous bases can be related with a discrete orthonormal 
basis labeled by the natural numbers via the Hermite functions. In consequence, we have continuous and discrete bases within the same framework. However, only discrete bases as complete orthonormal sets have a precise meaning in Hilbert space. If we have a structure allowing to work with these types of bases and to find relations among them, one needs to extend the Hilbert space to a more general structure called the rigged Hilbert space. 

The fundamental message of the present paper is to show how a class of different and apparently unrelated mathematical objects, such as classical orthogonal polynomials, Lie algebras, Fourier analysis, continuous and discrete bases and rigged Hilbert spaces, can be fully wived as a branch of harmonic analysis, with applications in quantum mechanics and signal processing, among other possible applications.

We have mentioned that the mathematical concept of rigged Hilbert space is very important in our work, although we shall not particularly stress this role here as we are mainly interested on the implications of the above mentioned tools in Fourier analysis. As rigged Hilbert spaces may not be familiar to many readers, let us briefly recall on this concept. It has been introduced by Gelfand and collaborators \cite{GEL} in connection with the spectral theory of self-adjoint operators. They proved, together with Maurin \cite{MAU}, the nuclear spectral theorem as was heuristically introduced by Dirac \cite{dirac}.

In fact, a rigged Hilbert space or Gelfand triple is a set of three vector spaces
\[\label{1}
\Phi\subset{\mathcal H}\subset \Phi^\times\,,
\]
where $\mathcal H$ is an infinite dimensional separable (i.e., with a countable orthonormal basis) Hilbert space, $\Phi$ is a topological vector space endowed with a topology finer than the Hilbert space topology  and dense on $\mathcal H$ with the Hilbert space topology, and $\Phi^\times$ is the dual space of $\Phi$ (i.e., the space of linear (or antilinear) continuous mappings from $\Phi$ into the set of complex numbers $\mathbb C$) and it is endowed with a topology compatible with the dual pair $(\Phi,\Phi^\times)$.

The rigged Hilbert space  formulation of quantum mechanics was introduced by Bohm and Roberts in the sixties of the last century and further developed later \cite{RO,ANT,B,MEL,GG,GG1,GG2}. Continuous bases are well defined in the dual $\Phi^\times$. The action of a functional $F\in\Phi^\times$ on a vector 
$\varphi\in\Phi$ is usually written as $\langle\varphi|F\rangle$ in order to keep up with the Dirac notation.  We shall  assume the antilinearity of the elements in $\Phi^\times$ because  we consider the scalar product on Hilbert space is antilinear to the left.

The first part of the present paper is devoted to a review of a previous work by the authors \cite{CGO16} concerning to the above mentioned extension of Fourier analysis on the real line with the use of special functions such us Hermite functions, which will be here our main example. This is studied in Section 2 and with the help of the Fractional Fourier transform in Section 3. 

In addition, we give a second example in which the real line has been replaced by the semi-axis ${\mathbb R}^+\equiv [0,\infty)$ and Hermite functions by Laguerre functions. In this latter case, we can construct two Fourier-like transforms  ${\mathcal T}^\pm$ with appropriate eigenvectors defined as functions on the half-line. This is given in Sections 4 and 5. Extensions to ${\mathbb R}^n$ using or not spherical coordinates are also possible, although we shall not consider this option in the present manuscript \cite{CGO18}. In Ref.  \cite{CGO17} we revisited  the harmonic analysis on the  group $SO(2)$ using the framework of  the rigged Hilbert spaces and in Ref. \cite{CGO18sp} a new realization of the group $SU(2)$ in the plane in terms of the associated Laguerre polynomials has been introduced.

Moreover, we introduce some new results concerning harmonic analysis on the circle. We construct new functions on the circle using Hermite functions and taking advantage of their properties. Again, these new functions give a unitary view of different mathematical objects that are often considered as unrelated: Fourier transform, discrete Fourier transform, Hermite functions and rigged Hilbert spaces. 

To understand the importance of the present research, let us remark that that Hermite and Laguerre functions are bases of spaces of square integrable functions, no matter whether real or complex, defined on $\R$ and $\R^+$, respectively. Square integrable real and complex (wave) functions play a similar role in signal processing and quantum mechanics, respectively. Besides, the interest of signal processing comes after the definition of two new types of filters. The former is based in restrictions to subspaces of $L^2(\R)$  or $L^2(\R^+)$. We have systematically constructed these filters by the use of the fractional Fourier transform. The second one is to choose low values of the index $n$ in the span of a given function by either Hermite or Laguerre functions (we may also use a combination thereof). These filters clean the signal or the wave function from noise or other unwanted effects. 
  
In addition, since the basic  operators related with these functions span some Lie algebras, such as the $io(2)$ \cite{GO91} for the Hermite functions and the $su(1,1)$ for the Laguerre functions, we can introduce a richer space of operators on $L^2(\R)$ or $L^2(\R^+)$, related to the universal enveloping algebra of $io(2)$ or $su(1,1)$, respectively\cite{EM}.

 These operator spaces connect functions describing the formal time evolution from a state to another in terms of filters or some kind of interaction.

\section{Harmonic analysis on ${\R}$}\label{hermite}

The first example of Fourier analysis and its relation with group theory is the case of the  one-dimensional translation group $T_1\simeq \R$ (a similar version but related to  the group $SO(2)$ can be found in \cite{CGO17}).  Its  unitary irreducible representations, $\mathcal R$, on the  continuous basis  
$\{|p\rangle\}_{p\in{\R}}$ of eigenvectors of the  infinitesimal generator ${P}$ of the group  is given by
\be\label{2.0}
 {\mathcal R}(x)|p\rangle=e^{-iPx}\,|p\rangle=e^{-ipx}\,|p\rangle\,;\qquad {P}\,|p\rangle=p\,|p\rangle\,,\quad \forall p\in \R\,, \forall x\in T_1\,.
 \ee
 The vectors of the   basis $\{|p\rangle\}$ verify
\be\label{3}
\langle p|p'\rangle = \sqrt{2\pi}\,\delta(p-p')\,,\qquad 
\frac 1{\sqrt{2\pi}} \int_{-\infty}^\infty |p\rangle\langle p|\,dp={\I}\,.
\ee
Considering the position operator $X$ and a continuous basis  $\{|x\rangle\}_{x\in{\R}}$ of eigenvectors of it, i.e.
\be\label{position}
 {X}\,|x\rangle=x\,|x\rangle\,,\qquad  \forall x\in \R\simeq T_1\,.
 \ee
Via  the Fourier transform  we can relate both (conjugate) bases $\{|p\rangle\}$ and $\{|x\rangle\}$
\be\label{xpfourier}
|x\rangle = \frac 1{\sqrt{2\pi}} \int_{-\infty}^\infty\, e^{-ipx}\, |p\rangle\, dp, \qquad
|p\rangle = \frac 1{\sqrt{2\pi}} \int_{-\infty}^\infty\, e^{ipx}\, |x\rangle\, dx, 
\ee
such that we find for the basis $\{|x\rangle\}$ that
\be
\langle x|x'\rangle= \sqrt{2\pi}\,\delta(x-x')\,, \qquad
  \frac 1{\sqrt{2\pi}} \int_{-\infty}^\infty |x\rangle\langle x|\,dx={\I}\,.
\ee
Moreover the operators ${X}$, ${P}$ and $\I$ close the Weyl-Heisenberg algebra
\be
[X,P]= i\,\I\,,\qquad [\cdot,\I]=0\,.
\ee
For more details see Ref~\cite{wu}.


\subsection{Hermite functions and the group $IO(2)$}\label{ios2group}

Now we can consider the inhomogeneous orthogonal group $IO(2)$ which is isomorphic to the Euclidean group in the plane, $E(2)$. For the study of the projective representations \cite{HAM,BAR,CT} we have to deal with the central extended group \cite{GO91}.  We take into account  a non-standard approach related to the ray representations of the inhomogeneous orthogonal group $IO(2)$ by considering the algebra of the harmonic oscillator that it is isomorphic to the central extension mentioned above. So, we have to consider the operators
 \[
 a:=\frac{1}{\sqrt{2}}\,(X+iP)\,,\qquad
 a^+:=\frac{1}{\sqrt{2}}\,(X-iP)\,,\qquad
 N:=a\,a^+\,,\qquad \I\,,
\]
which determine the Lie commutators
\[
[a,a^+]=\I\,,\qquad
[N,a]=-a\,,\qquad
[N,a^+]=a^+\,,\qquad
[\I,\cdot]=0\,,
\]
and the quadratic Casimir
\[
{\cal C}=\{a,a^+\}-2(N+1/2)\,\I\,.
\]
In the representation for ${\cal C}=0$ we obtain the differential equation
\be\label{DE}
{\cal C}\,K_n(x)\equiv \left(-D_x^2+X^2-(2N+1)\right)\,K_n(x)=0\,,
\ee
where $P=-i\,D_X=-i\,d/dx$ and $N$ is an operator acting in the index $n\in \N$ of the  functions
$K_n(x)$ solutions of the differential equations \eqref{DE}, i.e. $N\,K_n(x)=n\,K_n(x),\, n\in \N$. 
These solutions are the Hermite functions
 \begin{equation}\label{5}
K_n(x):=\frac{e^{-x^2/2}}{\sqrt{2^nn!\sqrt\pi}}
\;H_n(x)\,,
\end{equation}
where $H_n(x)$ are the Hermite polynomials, that determine an orthonormal basis in $L^2({\mathbb R})$
\begin{equation}\label{6}
\int_{-\infty}^\infty K_n(x)\,K_m(x)\,dx=\delta_{nm}\,,\qquad
\sum_{n=0}^\infty K_n(x)\,K_n(x')=\delta(x-x')\,.
\end{equation}
Note that we denote by $\N$ the set of positive integers or natural numbers together with 0 and by $\N^*=\N-\{0\}$. 

We see that the spectrum of the operator $N$ is infinite but countable and  we are able to construct a countable orthonormal basis of eigenvectors of $N$, $\{|n\rangle\}_{n\in\N}$, in terms of the continuous basis related to $X$ and the Hermite functions 
\begin{equation}\label{7}
|n\rangle:= (2\pi)^{-1/4}\int_{-\infty}^\infty K_n(x)\,|x\rangle\,dx\,,\qquad n=0,1,2,\dots\, .
\end{equation}
From the properties of the continuous basis as well as of the Hermite functions we obtain that
\begin{equation}\label{8}
\langle n|m\rangle=\delta_{nm}\,,\qquad \sum_{n=0}^\infty |n\rangle\langle n|={\I}\,.
\end{equation}

It is worthy noticing that the Hermite functions are eigenfunctions of the Fourier transform 
\begin{equation}\label{9}
[{\mathcal F}\,K_n](p)\equiv \widetilde K_n(p)
=\frac 1{\sqrt{2\pi}}  \int_{-\infty}^\infty e^{ipx}\,K_n(x)\,dx
=i^n\,K_n(p)\,,
\end{equation}
which allows us to related the three bases that we have defined in this section. 
The new ones are
\begin{eqnarray}
  |n\rangle &=& i^n (2\pi)^{-1/4}\int_{-\infty}^\infty K_n(p)\,|p\rangle\,dp \,,\label{11}\\[2ex]  |x\rangle &=& (2\pi)^{1/4}\sum_{n=0}^\infty K_n(x)\,|n\rangle \,, \label{12}\\[2ex]
   |p\rangle &=& (2\pi)^{1/4} \sum_{n=0}^\infty i^n\,K_n(p)\,|n \rangle\,.\label{13}
\end{eqnarray}
We see that the Hermite functions are the elements of the ``transition matrices'' between the continuous ad the discrete  bases.
We can express any ket $|f\rangle$ in any of the three bases obtaining the expressions
\be\label{14}
|f\rangle= \frac1{\sqrt{2\pi}} \int_{-\infty}^\infty dx\,f(x)\,|x\rangle \,,\qquad
|f\rangle = \frac1{\sqrt{2\pi}}  \int_{-\infty}^\infty dp\,  \tilde f(p)^*\,|p\rangle
=
(2\pi)^{-1/4}\,\sum_{n=0}^\infty a_n\,|n\rangle\,,
\ee
with
\be\label{16}
f(x):=\langle x|f\rangle= \sum_{n=0}^\infty \,a_n\,K_n(x)\,,\qquad
\tilde f(p)^*:= \langle p|f\rangle= \sum_{n=0}^\infty (-i)^n\,a_n\,K_n(p)\,,
\ee
and
\be\label{18}
a_n=
(2\pi)^{1/4}\, \langle n|f\rangle =     \int_{-\infty}^\infty dx\,K_n(x)\,f(x)=
i^n \, (2\pi)^{-1/4} \int_{-\infty}^\infty  dp\,K_n(p)\,\tilde f(p)^*\,.
\ee
So, we have obtained three different ways for expressing a quantum state $|f\rangle$ in terms of  three different bases: two continuous and non-countable $\{|x\rangle\}_{x\in\R}$, $\{|p\rangle\}_{p\in\R}$ and $\{|n\rangle\}_{n\in\N}$  infinite but countable. Nevertheless the framework to deal together with all these three bases is a rigged  Hilbert space \cite{B}. 

In particular,  the set $\{ |n\rangle\equiv K_n(x)\}_{n\in\N}$ is a discrete basis of $\Phi\equiv {\cal S}$ (the Schwartz space) and 
${\mathcal H\equiv L^2(\R)}$ and the continuous bases belong to $\Phi^\times\equiv {\cal S}^\times$ (the space of tempered distributions). More  precise we have two equivalent rigged Hilbert spaces: an abstract one  $\Phi\subset{\mathcal H}\subset \Phi^\times$ and other one, realized in terms of functions, 
$\mathcal S \subset L^2({\mathbb R}) \subset \mathcal S^\times$ through the unitary map  $U:\mathcal H\longmapsto L^2({\mathbb R})$ defined by $U |n\rangle= K_n(x)$.
Another interesting fact related with the framework of the rigged Hilbert spaces is that 
the space  $\mathcal S$ belongs to the domain of all operators in the universal enveloping algebra of $iso(2)$, $UEA[io(2)]$, and also these operators  can be extended by duality to operators on ${\mathcal S}^\times$, which are continuous under any topology on $\mathcal S^\times$ compatible with the dual pair. For a detailed exposition of  the actual case see \cite{CGO16} and references therein.

 \subsection{$UEA[io(2)]$ and fractional Fourier transform }\label{substructuresUEA}

Let us consider the kets $|n\rangle$ that constitute a complete orthonormal system in the abstract Hilbert space 
$\mathcal H$. For any  $n\in \N$ and $0<k\leq n\in \N$ we can consider the natural numbers, $q$ and $r$ such that
$n=k\, q+r\,,$
 where   $r=0,1,2,\dots,k-1$. For $k$ fixed the set  $\{|k\,q+r\rangle\}$ is a complete orthonormal system in  $\mathcal H$. Let us define the  operators 
 \begin{equation}\begin{array}{l}\label{47}
Q\,|k \,q+r\rangle :=q\,|k  \,q+r\rangle\,,\qquad
R\,|k  \,q+r\rangle :=r\,|k \,q+r\rangle\,.
\end{array}\end{equation}
These operators also act on $\Phi\subset {\mathcal H}$ and can be extended by duality to $\Phi^\times$.

The universal enveloping algebra $UEA[io(2)]$   contains infinitely many copies of the Lie algebra $io(2)$. Thus, for any positive integer $k$ we have the pairs $(k,r)$ with $0\leq r\leq k-1$ such that each pair  labels a copy of $io(2)$ denoted by $io_{k,r}(2)$
and 
\[\label{46}
\bigoplus_{r=0}^{k-1}\; io_{k,r}(2) \subset UEA[io(2)]\,.
\]
We define the family of operators $A_{k,r}^\dagger$ and $A_{k,r}$ belonging to  $UEA[io(2)]$ by
\[
A_{k,r}^\dagger := (a^\dagger)^k  \frac{\sqrt{N+k-r}}{\sqrt{k \prod_{j=1}^{k}(N+j)}}\,\,,\qquad  
A_{k,r} := \frac{\sqrt{N+k-r}}{\sqrt{k \prod_{j=1}^{k}(N+j)}} (a)^k\,,
\]
where $A_{k,r}^\dagger$ is   the formal adjoint of $A_{k,r}$ and viceversa. They  are continuous on $\Phi$ and can be extended by continuity to the dual $\Phi^\times$. Their action on the  vectors $|k\,q+r\rangle$ is
\[\label{57}
A_{k,r}^\dagger\, |k\, q+r\rangle =  \sqrt{q+1}\; |k\,(q+1)+r\rangle \,, \qquad
A_{k,r}\, |k\, q+ r\rangle = \sqrt{q}\; |k\,(q-1)+r\rangle \,.
\]
For each pair of integers $k$ and $r$ with $0\le r < k$, the operators $ Q,A_{k,r}^\dagger,A_{k,r}$ and $\I$ close a $io(2)$ Lie algebra  denoted by $io_{k,r}(2)$ 
\[\begin{array}{lllrll}\label{58}
[Q,\, A_{k,r}^\dagger]&=& +\, A_{k,r}^\dagger\,,\qquad &[Q, A_{k,r}]&=& -\,
A_{k,r}\,, \\[0.4cm]
[ A_{k,r},\, A_{k,r}^\dagger]& =& \I \,, \qquad &[ \I ,\cdot] &=& 0\,.
\end{array}\]
Note  that for any pair  $(k,r)$  the kets $ |k\, q+r\rangle$  span subspaces ${\mathcal H}_{k,r}$ of ${\mathcal H}$ and  $L^2_{k,r}({\mathbb R})$ of $L^2({\mathbb R})$.  Hence, we have that
\begin{equation}\label{61}
{\mathcal H}=\bigoplus_{r=0}^{k-1} {\mathcal H}_{k,r}\,,\qquad L^2({\R})=\bigoplus_{r=0}^{k-1}
 L^2_{k,r}({\R})\,.
\end{equation}
On the other hand, the spaces $\Phi_{k,r}$ and $\mathcal S_{k,r}$ can be easily obtained. A vector $|\phi\rangle$ belongs to $\Phi_{k,r}$ if and only if
\begin{equation}\label{62}
|\phi\rangle=\sum_{q=0}^\infty a_q\,|k\,q+r\rangle\,,
\end{equation}
such that
\[\label{62bis}
\sum_{q=0}^\infty |a_q|^2\,(q+1)^{2p}<\infty\,, \quad p=0,1,2,\dots
\]
A similar result can be achieved for any function in $\mathcal S_{k,r}$, just replacing $|k\,q+r\rangle$ in (\ref{62})  by the normalized Hermite functions $K_{k\,q+r}(x)$. 
Moreover the corresponding  rigged Hilbert spaces can be obtained
\be\begin{array}{l}\label{2.32}
\Phi_{k,r}\subset {\mathcal H}_{k,r}\subset \Phi_{k,r}^\times\,,\\[0.4cm]
\mathcal S_{k,r} \subset L^2_{k,r}({\R}) \subset \mathcal S_{k,r}^\times\,.
\end{array}
\ee
One can also prove that an operator ${\mathcal O}$ belongs to  $UEA[io_{k,r}(2)]$ if and only if ${\mathcal O}$ is an operator of ${\mathcal H_{k,r}}$.

The split of $L^2({\mathbb R})$  as direct sum of subspaces  $L^2_{k,r}({\R})$ is connected with the fractional Fourier transform, which is is a generalization of the Fourier transform  \cite{OZA}. It is very interesting that we can also relate the 
 fractional Fourier transform with the Hermite functions $K_n(x)$ \eqref{5} as follows.
The {fractional Fourier transform} of $f\in L^2({\mathbb R})$ associated to $a\in\R$, ${\mathcal F}^af$, is given by
\begin{equation}\label{64}
[{\mathcal F}^a f](p):= \sum_{n=0}^\infty a_n\,e^{i\,n\,a\,\pi/2}\,K_n(p)\,,
\end{equation}
where
\begin{equation}\label{63}
f(x)=\sum_{n=0}^\infty a_n\,K_n(x)\,,\qquad
  a_n=\int_{-\infty}^\infty f^*(x)\,K_n(x)\,dx\,,
\ee
Series in (\ref{64}) obviously converges in both the $L^2({\mathbb R})$ norm and in the sense given in (\ref{62}) provided that $f(x)\in\mathcal S$, so that ${\mathcal F}^a f\in\mathcal S$ if $f\in\mathcal S$.

If we consider the case $a=4/k$, where $k$ is a positive integer,  we get
\be\label{2.34}
\tilde f^k (p):= [{\mathcal F}^{4/k}\,f](p)
=
\sum_{n=0}^\infty a_n\,e^{2\, \pi\,i\,n/k}\,K_n(p)\,.
\ee
Note that we recover the standard  Fourier transform when $k=4$, i.e. $a=1$.
 Since for every $k\in \N ^*$ any  $n\in \N$ can be decomposed as $n=k\,q+r$ with $q\in \N$ and  $0\leq r\leq k-1\,,$  we have the following decomposition of   $\tilde f^k$ as
\be\begin{array}{lll}\label{65}
\tilde f^k(p)&=&\displaystyle
\sum_{q=0}^\infty a_{kq} \,e^{2\pi (kq)i/k}\,K_{kq}(p) + \sum_{q=0}^\infty a_{kq+1}
 \,e^{2\pi (kq+1)i/k}\,K_{kq+1}(p)
\\[0.4cm]
&& \displaystyle
\qquad+ \cdots + \sum_{q=0}^\infty a_{kq+k-1} \,e^{2\pi (kq+k-1)i/k}\,K_{kq+k-1}(p) \\[0.4cm]
&=&   \tilde f_0^k(p)+\tilde f_1^k(p)+\dots +\tilde f_{k-1}^k(p)\,,
\end{array}\ee
where
\[
f_r^k(x):= \sum_{q=0}^\infty a_{kq+r} \, K_{kq+r}(x)
\,,\qquad
 \tilde f_r^k(p):=e^{2\,\pi \,r\,i/k}\, f_r^k(p)\,.
\]
The vectors $\tilde f^k_r$,\, ($r=0,1,\dots,k-1$)\, are mutually orthogonal, so that (\ref{65}) gives a split of  $L^2({\R})$ into an orthonormal direct sum of subspaces. Furthermore, having in mind that
$
\tilde f^k_r(p) \equiv
[{\mathcal F}^{4/k}\,f^k_r](p) $,
we see that each term in the direct sum is an eigen-subspace of ${\mathcal F}^{4/k}$ with eigenvalue $e^{i 2\pi r/k}$, i.e., independently from the bases $\{|x\rangle\}$ or $\{|p\rangle\}$
\[\label{66}
L^2({\mathbb R})=L^2_{k,0}({\R}) \oplus L^2_{k,1}({\R}) \oplus \dots \oplus L^2_{k,k-1}({\R})\,.
\]
 Thus, we have recovered the decomposition displayed in \eqref{62} and \eqref{2.32}

\section{Harmonic analysis  on ${\R}^+$}\label{fourierlikermas}
\label{laguerre}

Let $L^2({\mathbb R}^+)$ be the space of square integrable functions on ${\R}^+ \equiv [0,+\infty)$. 
As is well known a  complete orthonormal set in $L^2({\mathbb R}^+)$ is determined by the functions
\begin{equation}\label{68}
M_n^\alpha(y) = \sqrt{\frac{\Gamma(n+1)}
{\Gamma(n+\alpha+1)}}\,y^{\alpha/2}\,e^{-y/2}
\,L_n^\alpha(y)\,,
\end{equation}
where $-1<\alpha<+\infty$, $n = 0,1,2,\dots\,,$   and $L_n^\alpha(y)$ are the associated Laguerre polynomials \cite{AS72,OLBC}.  The Laguerre functions, $M_n^\alpha(y)$,  verify 
the following  orthonormality and completeness relations
\begin{equation}\label{75}
 \int_{0}^{\infty} M_{n}^{\alpha}(y)\, \, M_{m}^{\alpha}(y) \;dy\;=\; \delta_{n m}\,\,,\qquad 
 \displaystyle \sum_{n=0}^{\infty}  \;M_{n}^{\alpha}(y)\,\, M_{n}^{\alpha}(y')\; =\;
\delta(y-y')\,.
\end{equation}

 \subsection{Harmonic analysis on $su(1,1)$}\label{su11group}

Considering the  operators on $L^2(\R^+)$
\be\begin{array}{lllll}\label{3.5}
Y\, M_n^\alpha(y)& :=& y\, M_n^\alpha(y)\,,
\qquad  D_y\, M_n^\alpha(y)&:= &\ds M_n^\alpha(y)'=\frac{d}{dy}M_n^\alpha(y)\,, \\[0.4cm]
{N}\,  M_n^\alpha(y) &:= & n\,  M_n^\alpha(y)\, , \qquad
{\I}\,  M_n^\alpha(y)& := &   M_n^\alpha(y)\,,
\end{array}\ee
we can construct the operators
\be\label{3.6}
J_+ := \left(Y\, D_y +N+1 +\frac{\alpha-Y}{2}\, \right)\,, \qquad
J_-\;  \left( -Y\, D_y +N +\frac{\alpha-Y}{2}\right)  \,.
\ee
They act  on the basis functions $M_n^\alpha(y)$ as
\be  \label{3.8} \begin{array}{l}
J_+ M_n^\alpha(y)  = \sqrt{(n+1)(n+\alpha+1)} M_{n+1}^\alpha(y)\\[0.4cm]
J_- M_n^\alpha(y)   =
\sqrt{n(n+\alpha)} M_{n-1}^\alpha(y).
\end{array}\ee
These two operators together with 
\[
J_3:=N+\frac{\alpha+1}{2}\,\I\,, \qquad
J_3\, M_n^\alpha(y)=\left(n+\frac{\alpha +1}{2}\right)\, M_n^\alpha(y)\,,
\]
close the Lie algebra $su(1,1)$ because their 
commutation  relations  are
\begin{equation}\label{90}
[J_3, J_\pm] = \pm J_\pm \,,\qquad [J_+, J_-] =  - 2\, J_3\,.
\end{equation}
The Casimir operator $\mathcal C$ of  $su(1,1)$ is
\begin{equation}\label{94}
{\mathcal C}={J}_3^{\,2}-\frac 12 \{{J}_+,{J}_-\}  =
 \frac{\alpha^2-1}{4}\,\I\,.
\end{equation}
From (\ref{3.6}) we obtain that
$Y=-(J_++J_-)+2N+(\alpha+1)\,\mathbb I\,,$ and from the Casimir it is possible to obtain the differential equation defining the associated Laguerre polynomials.

Omitting the technical details that the interested reader can find in Ref \cite{CGO16}, there exists a set of generalized eigenvectors of $Y$,  $\{|y\rangle\}_{y \in \mathbb R^+}$,  (or more strictly of  $U^{-1}YU$, where  $U:\mathcal H\longmapsto L^2(\mathbb R^+)$ is a unitary operator and $\mathcal H$ be an arbitrary infinite dimensional separable Hilbert space) such that
\be\label{76}
{Y}|y\rangle= y|y\rangle\,,\;\qquad \langle y|y'\rangle=\delta(y-y')\,, \qquad
 \int_{-\infty}^\infty |y\rangle\langle y|\,dy=\mathbb I\,.
\ee
Actually we have two (families, depending on $\alpha$,  of) equivalent rigged Hilbert spaces  $\Phi_\a\subset{\mathcal H}\subset \Phi_\a^\times$  and  ${\mathcal D}_\a\subset L^2(\mathbb R^+)\subset {\mathcal D}_\a^\times$. All the elements  and their extensions of the $UEA(su(1,1)$ are continuous  on both rigged Hilbert spaces.
 
In analogy with the case of the whole real line  a decomposition like (\ref{61}) for any $k\neq 0\in N$ is also obtained
\begin{equation}\label{100}
{\mathcal H}=\bigoplus_{r=0}^{k-1} {\mathcal H}_{k,r}\,,\quad L^2({\mathbb R}^+)=\bigoplus_{r=0}^{k-1} L^2_{k,r}({\mathbb R}^+)\,.
\end{equation}

For any  $n\in {\N}$ and $\alpha\in (-1,+\infty)$, we can  define the vectors $|n,\a\rangle\in\Phi_\a$ as
\begin{equation}\label{77}
|n, \a\rangle:= \int_0^\infty dy\, M_n^\alpha(y)|y\rangle\,, 
\end{equation}
that from  (\ref{75}) and (\ref{76}) have the properties
\begin{equation}\label{78}
\langle n,\a|m,\a\rangle=\delta_{nm} \,, \qquad \sum_{n=0}^\infty |n,\a\rangle\langle n,\a|=\mathbb I\,.
\end{equation}
Hence, $|n,\a\rangle$ with $n\in {\N}$ (and $\alpha$  fix)  is a complete orthonormal set (orthonormal basis) in $\mathcal H$.  Taking into account the unitary operator $U$ we have $U|n,\a\rangle=M_n^\alpha$. For $y\ge 0$, we easily obtain  that
\begin{equation}\label{79}
\langle y|n,\a \rangle =
\int_0^\infty dy'\,M_n^\alpha(y')\,\langle y|y'\rangle
 =M_n^\alpha(y)\,.
\end{equation}
Note that as we see for the whole real line we have  also obtained  two different bases for the elements in $\Phi_\a\subset {\mathcal H}$: a continuous basis
$\{|y\rangle\}_{y\in\R^+}$ and a discrete basis $\{|n,\a\rangle\}_{n\in \N}$ whose elements are eigenvectors of the operator
$U^{-1}NU$  where $N$ was defined in \eqref{3.5}.


\subsection{Fourier-like transformations  on ${\R}^+$}
\label{laguerrefourier}

In section \ref{substructuresUEA}, we have presented the fractional Fourier transform associated to the Hermite functions. Now after the results displayed in the previous section that show a close  analogy between  the formalisms on $\R$ and on $R^+$ we can have the temptation to extend the Fourier fractional  formalism to the generalized Laguerre functions. Nevertheless, this is not possible since the Laguerre functions, $M_n^\alpha(y)$, are not eigenfunctions of the Fourier transform unlike the Hermite functions. 
Fortunately there exists a partial way out because the relations between Hermite and Laguerre polynomials
\[\label{101}
H_{2n}(x) = (-1)^n \,2^{2n}\, n! \,L^{-1/2}_n(x^2) \,,\qquad
H_{2n+1}(x) =  (-1)^n\, 2^{2n+1}\, n! \, x\, L^{+1/2}_n(x^2)\,,
\]
that give equivalent relations between the Hermite and the Laguerre functions
\[
K_{2n}(x)=(-1)^n \,(x^2)^{1/4}\,  \,M^{-1/2}_n(x^2)\,,\qquad
K_{2n+1}(x)=(-1)^n \,x\,(x^2)^{-1/4}\,  \,M^{1/2}_n(x^2)\,.
\]
Defining the transforms ${\mathcal T}_\pm$  on functions $f(y)\in L^2({\mathbb R}^+)$ by
\be
[{\mathcal T}_\pm f](s):=  \frac{1}{\sqrt{2 \pi}} \int_0^\infty dy\, \frac{S_\pm (\sqrt{s\, y})}{(s\, y)^{1/4}}\,f(y)\,, \label{103} 
\qquad
S_+(\cdot)=\cos (\cdot),\;\; S_-(\cdot)=\sin (\cdot)
\ee
we can see that they verify 
\begin{equation}\label{105}
[{\mathcal T}_\pm M_n^{\pm 1/2}](s)=(-1)^n\,M_n^{\pm 1/2}(s)\,,
\end{equation}
i.e.,  $M_n^{\pm 1/2}(s)$ are eigenfunctions of the operators ${\mathcal T}_\pm$, respectively, with eigenvalues $(-1)^n$. Hence we have two relevant values of  the label $\a$: $\pm 1/2$.  Then,  for any $f(y)\in L^2({\mathbb R}^+)$ since
\be\label{106}
f(y)=\sum_{n=0}^\infty a_n^\pm\,M_n^{\pm1/2}(y)\,,
\qquad 
a_n^\pm=\int_0^\infty f^*(y) M_n^{\pm1/2}(y)\,dy\,,
\ee
we can define two new   fractional integral transforms for each $a\in \R$, ${\mathcal T}_\pm ^a$, by
\[
[{\mathcal T}_\pm ^a\,f](s):=
\sum_{n=0}^\infty a_n^\pm\,e^{i\,n\,a\,\pi/2}\,M_n^{\pm1/2}(s)\,.
\]
In the case of $a=4/k$ with $k\in \N^*$ we have
\begin{eqnarray}
\tilde f_\pm^k (s)&:=&[{\mathcal T}_\pm ^{4/k} f](s)= \sum_{q=0}^\infty a_{kq}^\pm\, e^{-2\pi(kq)i/k}\,M_{kq}^{\pm 1/2}(s) + \sum_{q=0}^\infty a_{kq+1}^\pm\, e^{-2\pi(kq+1)i/k}\,M_{kq+1}^{\pm 1/2}(s)  +\dots \nonumber\\[2ex]
&&\qquad\qquad\qquad  \dots +\sum_{q=0}^\infty a_{kq+k-1}^\pm\,e^{-2\pi(kq+k-1)i/k}\,M_{kq+k-1}^{\pm 1/2}(s) \nonumber\\[2ex]
&=& f_{0,\pm}^k (s)+ e^{-2\pi i/k}  \,f_{1,\pm}^k(s)+\dots + e^{-2\pi(k-1)i/k}\, f_{k-1, \pm}^k (s)\,, \label{107}
\end{eqnarray}
with
\[
f_{r,\pm}^k (s):=\sum_{q=0}^\infty a_{kq+r}^\pm\, M_{kq+r}^{\pm 1/2}(s)\,.
\]
 We have recover the  decomposition (\ref{100}) of $L^2({\mathbb R}^+)$ for the particular cases of $\a=\pm 1/2$
\[\begin{array}{l}\label{108}
L^2({\mathbb R}^+) = L_{k,0}^2({\mathbb R}^+)^\pm \oplus L_{k,1}^2({\mathbb R}^+)^\pm \oplus\dots\oplus L_{k,k-1}^2({\mathbb R}^+)^\pm\,,
\end{array}\]
where each of the closed subspaces $L_{k,r}^2({\mathbb R}^+)^\pm$ is an eigen-subspace of ${\mathcal T}_\pm$ with eigenvalue $e^{-2\,\pi \,r\,i/k}$.  

\section{A new harmonic analysis on the circle}\label{circlehermite}

The Hermite functions $K_n(x)$ allow us to construct  a countable set of periodic functions that constitute a system of generators of the space of square integrable functions on the unit circle $L^2(\mathcal C)$, i.e.,  
 the functions $f(\phi):\mathcal C\longmapsto \mathbb C$ with norm $\big|\big| f(\phi)\big|\big|$ defined by
 \begin{equation}\label{2}
 ||f(\phi)||^2:=\frac{1}{2\pi}\,\int_{-\pi}^{\pi}|f(\phi)|^2\,d\theta<\infty\,.
\end{equation}
Let us define the  periodic  functions (with period $2\pi$)
\begin{equation}\label{3.0}
{\mathcal K}_n(\phi):= \sum_{k=-\infty}^\infty K_n(\phi+2k\pi)\,,\qquad   -\pi\le \phi<\pi\,, \;\;
n=0,1,2,\dots\,.
\end{equation}
It can be proved that the series defining the $\K_n(\phi)$ are absolutely convergent and also that every $\K_n(\phi)$ is bounded and square integrable on the interval $-\pi\le\phi<\pi$.  
From this property can be also proved using the Lebesgue theorem that
\begin{eqnarray}\label{11.0}
\int_{-\pi}^{\pi} e^{im\phi}\,d\phi \sum_{k=-\infty}^\infty K_n(\phi+2k\pi) =\sum_{k=-\infty}^\infty \int_{-\pi}^{\pi} e^{im\phi}\, K_n(\phi+2k\pi)\,d\phi\,.
\end{eqnarray} 

\subsection{A discretized Fourier transform on the circle}\label{circlefourier}
Let us compare the space $L^2(\mathcal C)$, also denoted as $L^2[-\pi,\pi]$, to the space $l_2(\mathbb Z)$ of 2-power summable sequences.   As is well known, an orthonormal basis on $L^2(\mathcal C)$ is $\{(2\pi)^{-1}\,e^{in\phi}\}$ with $n\in\mathbb Z$. Hence, any $f(\phi)\in L^2(\mathcal C)$ admits the following span (Fourier series)
\begin{equation}\label{12.0}
f(\phi)=\frac1{2\pi} \,\sum_{n\in\mathbb Z}f_n\,e^{in\phi}\,,\qquad f_n\in \C\,,
\end{equation}
with
\begin{equation}\label{12.00}
f_n=\frac1{2\pi} \,\int_{-\pi}^{\pi} f(\phi)\,e^{-in\phi}\,d\phi\,,
\end{equation}
and
where the sum \eqref{12.0} converges in the sense of the norm \eqref{2}. Moreover for continuous functions $f(\phi)$ the series also converge pointwise.  The properties of orthonormal basis in Hilbert spaces show that 
\begin{equation}\label{13.0}
\frac{1}{2\pi}\,\sum_{n\in\mathbb Z}|f_n|^2=  ||f(\phi)||^2\,.
\end{equation}
We may call to the sequence of complex numbers $\{f_n\}_{n\in\mathbb Z}$, the components of $f$.

The Hilbert space $l_2(\mathbb Z)$ is a space of sequences of complex numbers $A\equiv\{a_n\}_{n\in\mathbb Z}$ such that
\begin{equation}\label{14.0}
||A||^2:=\frac1{2\pi} \,\sum_{n\in\mathbb Z}|a_n|^2<\infty\,,
\end{equation}
 with scalar product given by
\begin{equation}\label{15}
\langle A|B\rangle:= \frac1{2\pi} \sum_{n\in\mathbb Z} a^*_n\,b_n\,.
\end{equation}
An orthonormal basis for $l_2(\mathbb Z)$ is given by the sequences  ${\mathcal E}_k=\{\delta_{k,n}\}_{n\in\Z}$ with $k\in \Z$.  Any $f\in l_2(\mathbb Z)$ with components  $\{f_n\}_{n\in\mathbb Z}$ may be written as
\begin{equation}\label{16.0}
f=\frac1{2\pi}\sum_{n\in\mathbb Z} f_n\,\e_n\,, \qquad 
\frac1{2\pi} \,\sum_{n\in\mathbb Z}|f_n|^2=||f||^2<+\infty\,.
\end{equation}
So, there exists a correspondence between $L^2(\mathcal C)$ and $l_2(\mathbb Z)$ relating any $f(\phi)\in L^2(\mathcal C)$ as in \eqref{12.0} with $f$ as in \eqref{16.0} with the same sequence $\{f_n\}_{n\in\mathbb Z}$. This correspondence, $f(\phi)\longmapsto f$, is clearly linear, one-to-one and onto. Also, $||f(\phi)||=||f||$, which shows that it is, in addition, unitary.

From expression \eqref{12.0} we see the expansion on   Fourier series of the functions of $L^2(\mathcal C)$. From this point of view, we may say that the Fourier series is a unitary mapping, $\mathcal F$, from $L^2(\mathcal C)$ onto $l_2(\mathbb Z)$. It admits an inverse, $\mathcal F^{-1}$,  from $l_2(\mathbb Z)$ onto $L^2(\mathcal C)$, which is also unitary and is sometimes called the discrete Fourier transform, i.e.
\be\label{discretefourier}
{\mathcal F}[f(\phi)]=\frac1{2\pi} \,\sum_{n\in\mathbb Z}f_n\,e^{in\phi}\equiv \{f_n\}_{n\in\N}\,,\qquad
{\mathcal F}^{-1}[\{a_n\}_{n\in\N}]=\frac1{2\pi} \,\sum_{n\in\mathbb Z}a_n\,e^{in\phi}\equiv a(\phi)\,,
\ee
with $f_n\in \C\,$ and given by \eqref{12.00}.

As we mention in the introduction we will give   a unitary version of concepts that are often introduced  separately, like  {Fourier transform}, {Fourier series} and {discrete Fourier transform} in one side and the {Hermite functions} on the other. 

We start by  constructing a set of functions in $l_2(\mathbb Z)$ using the Hermite functions $K_n(x)$.  We introduce the sequences $\chi_n$ associated to $K_n(x)$ as follows
\begin{equation}\label{17}
\chi_n:= \{K_n(m)\}_{m\in\mathbb Z}\,,\qquad n\in \N\,.
\end{equation}
These sequences $\chi_n$ are in $l_2(\mathbb Z)$. Moreover, they are linearly independent and span $l_2(\mathbb Z)$.
The proof can be find in \cite{CGO19} and they are based on the fact that 
\begin{equation}\label{18.0}
\left|\begin{array}{ccccc} H_0(-N) & \dots & H_0(0) & \dots & H_0(N) \\[2ex]
H_1(-N) & \dots & H_1(0) & \dots & H_1(N)   \\[2ex]
\dots & \dots & \dots & \dots & \dots  \\[2ex]
H_{2N}(-N) & \dots & H_{2N}(0) & \dots & H_{2N}(N) \end{array} \right| \ne 0\,,
\end{equation}
and
\begin{equation}\label{19}
\left| \begin{array}{cccc} H_0(0) & H_0(1) & \dots & H_0(N) \\[2ex] H_1(0) & H_1(1)  &\dots & H_1(N)
\\[2ex] \dots & \dots & \dots & \dots \\[2ex] H_{N}(0) & H_{N}(1) & \dots & H_{N}(N)
 \end{array} \right| \ne 0\,,
\end{equation}
for any $N\in \N$ where $H_n(k)$ is the Hermite polynomial $H_n(x)$ evaluated at the integer $k$ (remember that 
$ K_n(x)=e^{-x^2/2}\,H_n(x)/\sqrt{2^nn!\sqrt{\pi}}$). 

Since the functions $\K _n(\phi)$ are in $L^2[-\pi,\pi]$, they admit a span in terms of the orthonormal basis 
 $\{(2\pi)^{-1}\,e^{i m\phi}\}_{m\in\Z}$ in $L^2[-\pi,\pi]$. Thus, we can write
\begin{equation}\label{31}
\K_n(\phi)= \frac{1}{\sqrt{2\pi}} \sum_{m=-\infty}^\infty { k}_n^m\,e^{-im\phi}\,,
\end{equation}
with
\begin{equation}\label{32}
k_n^m= \frac{1}{\sqrt{2\pi}} \int_{-\pi}^\pi e^{im\phi}\,\K_n(\phi)\,d\phi\,.
\end{equation}
The continuity of the functions $\K_n(\phi)$ on $[-\pi,\pi]$ guarantees the pointwise convergence of \eqref{31} and since the $\K_n(\phi)$ are periodic with period $2\pi$, hence \eqref{31} is valid for all $\phi\in\R$. 

We recall that  the Hermite functions $K_n(x)$ are eigenfunctions of the Fourier transform with eigenvalue $(-i)^n$ \eqref{9}, i.e. $[\mathcal F\,K_n](p)= (-i)^n\,K_n(p)$. So $K_n(x)$ are eigenfunctions of the inverse Fourier transform with eigenvalue $i^n$, i.e. $[\mathcal F^{-1}\,K_n](x) =i^n\,K_n(x)$. From this fact we can  find an explicit expression of the coefficients $k_n^m$  \eqref{32} in terms of the values of the Hermite functions at the integers 
\[\begin{array}{lll}\label{34}
k_n^m &=&\ds \frac{1}{\sqrt{2\pi}} \int_{-\pi}^\pi e^{im\phi}\,d\phi \left[ \sum_{k=-\infty}^\infty K_n(\phi+2k\pi) \right] = \frac{1}{\sqrt{2\pi}} \sum_{k=-\infty}^\infty \int_{-\pi}^\pi e^{im\phi}\, K_n(\phi+2k\pi)\,d\phi \\[0.4cm] 
&=&\ds
 \frac{1}{\sqrt{2\pi}} \sum_{k=-\infty}^\infty \int_{-\pi+2k\pi}^{\pi+2k\pi}  e^{ims}\,K_n(s)\,ds = \frac{1}{\sqrt{2\pi}} \int_{-\infty}^\infty  e^{ims}\, K_n(s)\,ds = i^n\,K_n(m)\,,
\end{array}\]
were $s:=\phi+2k\pi$ and $e^{im\phi}=e^{im(\phi+2k\pi)}=e^{ims}$. Hence \eqref{31} and \eqref{32} can be written, respectively, as
\begin{equation}\label{35}
\K_n(\phi)= \frac{i^n}{\sqrt{2\pi}} \sum_{m=-\infty}^\infty K_n(m)\,e^{-im\phi}\,,
\qquad
K_n(m)= \frac{(-i)^n}{\sqrt{2\pi}} \int_{-\pi}^\pi  \K_n(\phi)\,e^{im\phi}\,d\phi\,,
\end{equation}
where $k_n^m={i^n}\, K_n(m)$.

The systems of generators in $L^2[-\pi,\pi]\equiv L^2(\mathcal C)$,  $\{\K_n(\phi)\}_{n\in \Z}$, and in  $l_2(\mathbb Z)$, the set of sequences $\{\chi_n\}_{n\in \Z}$, are not  orthonormal basis. 
The scalar product on $L^2[-\pi,\pi]$ is related with  the scalar product in $l_2(\mathbb Z)$ 
\[\begin{array}{ll}\label{38}
\langle \K_n|\K_m\rangle &= \ds
 \int_{-\pi}^\pi \K^*_n(\phi)\,\K_m(\phi)\,d\phi = \frac{1}{2\pi} \int_{-\pi}^\pi d\phi \sum_{j=-\infty}^\infty \sum_{k=-\infty}^\infty (-i)^n\,i^m \, K^*_n(j)\,K_m(k)\, e^{-i(k-j)\phi}  \\[0.4cm] 
 &=\ds  \sum_{j=-\infty}^\infty \sum_{k=-\infty}^\infty \delta_{j,k}\,i^{m-n}\,K_n^*(j)\,K_m(k) = \sum_{j=-\infty}^\infty  i^{m-n}\,K_n^*(j)\,K_m(j) \\[0.5cm] 
 &=\ds   i^{m-n}\,(\chi_n,\chi_m)\,.
\end{array}\]
The  Gramm-Schmidt procedure allows us to obtain orthogonal bases in both spaces. 

\section{Conclusions}

In this paper we have presented a unified framework where Hermite functions, or alternatively Laguerre functions, their symmetry groups, Fourier analysis and rigged Hilbert spaces fit in a perfect manner. Hermite functions play a fundamental role in the study of quantum mechanics and signal processing on the real line $\R$, while Laguerre functions play the same role on the half-line $\R^+$. We have also studied the particular relation between both situations.  In both cases, these functions are eigenvectors of the Fourier transform and this is an essential property.

It is precisely the use of rigged Hilbert spaces, what allows the use of discrete and continuous bases on a simple and interchangeable manner. This makes rigged Hilbert spaces the correct mathematical formulation that encompasses both quantum mechanics and signal processing. Here, Hermite functions act as transition elements of transition matrices between continuous as discrete bases. This is a quite interesting fact, which is important from the computational as well as the epistemological point of view. 

We have shown how Fourier analysis allows for the decomposition of rigged Hilbert spaces  into direct sums of rigged Hilbert spaces. This may permit the filtering of noise or any other undesirable addition. The same applies to operators as we may restrict their evolution to a sub-algebra, which has been chosen among infinite other possibilities in the universal enveloping algebra of the corresponding symmetry group. The decomposition of rigged Hilbert spaces is consistent with the fractional Fourier transform. This is the cornerstone of the filtering procedure. We have extended the formalism to functions over the semi-axis $\R^+$ by the construction of a pair of ``Fourier-like'' transformations which play the role before assigned to the Fourier transform on $\R$. Fractional Fourier transforms may be defined after these Fourier-like transforms and also serve for filtering. Moreover, the algebraic approach associated to the Lie symmetry algebra and its universal enveloping algebra allows us to extend all results from the vector spaces to the operators acting on them.

All these results can be also translated, in some sense, to the circle. We have constructed some special functions on the circle out of Hermite functions and have taken advantage of the properties of Hermite functions in order to use Fourier analysis on the circle as well.

As a final remark, let us insist that we have given a unitary point of view of mathematical objects that are often considered as unrelated such as Fourier transform, discrete Fourier transform, Hermite and Laguerre functions and rigged Hilbert spaces.

\section*{Acknowledgements}

This work was partially supported by the Ministerio de Econom\'ia y Competitividad of Spain (Project MTM2014-57129-C2-1-P) and the Junta de Castilla y Le\'on (Projects VA057U16 and VA137G18).


\end{document}